\documentclass[12pt]{article}

\def\ca{\c{c}\~{a}}

\def\ii{\'{\i}}

\newcommand{\ud}{\mathrm{d}}

\usepackage{color}

\def\mathbf{\vec}
\def\be{\begin{equation}}
\def\ee{\end{equation}}
\def\ba{\begin{eqnarray}}
\def\ea{\end{eqnarray}}

\def\bx
       {\hbox to.77778em{\hfil\vrule\vbox to.675em
       {\hrule width.6em\vfil\hrule}\vrule\hfil}}

\begin{document}


\centerline{\bf {\large Perturbative approach to $U_A(1)$ breaking\footnote {Prepared for the Proceedings of The International Conference on High Energy and Mathematical Physics, Marrakech, Morrocco, 3-7 April 2005}   }}

\vspace{0.5cm}

\centerline{Brigitte Hiller$^{(1)}$, Alexander A. Osipov$^{(1),(2)}$, Alex H. Blin$^{(1)}$}
\vspace{1cm}

{\small {\it (1) Centro de F\'{\i}sica Te\'{o}rica, Departamento de
         F\'{\i}sica da Universidade de Coimbra, 3004-516 Coimbra, Portugal
\vspace{.2cm}

(2) Joint Institute for Nuclear Research, Laboratory of Nuclear Problems, 141980 Dubna,
        Moscow Region, Russia}} 
\vspace{1cm}

\centerline{{\bf Abstract}}
\vspace{.5cm}

The six-quark instanton induced 
't Hooft interaction is considered in combination with the 
Nambu-Jona-Lasinio (NJL) type $U_L(3)\times U_R(3)$ chiral 
symmetric Lagrangian. We discuss the bosonization of this multi-quark interaction, taking the
$U_A(1)$ breaking as a perturbation. We discuss its relation with the usual approach.

\vspace{1cm}

{\bf Introduction}
\vspace{0.5cm}

In low-energy QCD phenomena one considers basic powerful classification schemes, as
symmetries and symmetry breaking patterns, large $N_c$ counting rules \cite{Hooft:1974}- \cite{Pich:2002} and
scales at which some type of interactions set in \cite{Shifman:1981}. 
A well known example is $\eta'$ meson physics, where all these concepts are needed to
explain its large mass $m_{\eta'}\sim 1\, GeV$. 
If  the singlet $SU(3)$ axial current of QCD, $U_A(1)$, were conserved as 
$m_{curr}\rightarrow 0$, the $\eta'$ would occur as the ninth pseudoscalar Goldstone. 
As the global $U_L(3)\times U_R(3)$ chiral symmetry of QCD is broken by 
the $U_A(1)$ Adler-Bell-Jackiw anomaly \cite{Adler:1969} , it was realized through the study of instantons \cite{Hooft:1976}-\cite{Hooft:1978} that
effective $2N_f$ quark vertices arise, the 't Hooft interactions , responsible 
for the non-conservation of the singlet axial current and absence of the 
related Goldstone boson. Typically, the mass squared of an approximate Goldstone boson is
linear in the symmetry breaking parameter. The anomaly is of order
$1/N_c$, one expects $m^2_{\eta'}$ to be of order $(1/N_c)$. 
Although the $U_L(3)\times U_R(3)$ symmetry is recovered in the limit 
$N_c\rightarrow\infty$, the empirical mass of the $\eta'$ is abnormaly 
large to conform alone with the large $N_c$ counting rules, which is Witten's puzzle \cite{Witten:1979a}.
The solution was given in \cite{Shifman:1981}, where with the help of QCD sum rules it was shown that a large scale emerges in the $0^-,0^+$ gluonic channels, compared to which the $\eta'$ mass is small, 
and characterizing the breaking of asymptotic feedom in these channels.

Our plan consists in
addressing the $\eta'$ physics and lowest lying ($0^-,0^+$) meson
nonets in a many-fermion vertices model without explicit gluon 
degrees of freedom. In this way one takes instead the quark structure of the mesons explicitly into account, and may obtain relevant information about the vacuum structure of QCD, albeit in an effective field theoretical approach.
We use the successful Nambu - Jona-Lasinio model \cite{Nambu:1961} extended to the $U(3)_L\times U(3)_R$
chiral symmetry of massless QCD and the six quark 't Hooft interactions for the 
$U_A(1)$ symmetry breaking.
We review some known bosonization schemes with functional integral methods and show their restrictions. Within these schemes the leading order $\eta'$ mass is obtained close to its 
empirical value and a good description of the remaining members of the 
lowest pseudoscalar nonet and related weak decay constants and quark 
condensates. The $\eta'$ puzzle is also present in this case.
Finally, using the spectral decomposition approach, we indicate how to obtain systematically all $U_A(1)$ corrections in a perturbative way and its relation to the previous approach.  
\vspace{0.5cm}

{\bf The model Lagrangian}
\vspace{0.5cm}

The starting Lagrangian is
\begin{equation}
\label{totlag}
  {\cal L}=\bar{q}(i\gamma^\mu\partial_\mu -\hat{m})q
          +{\cal L}_{NJL}+{\cal L}_6,
\end{equation}
with NJL four-fermion vertices of the scalar and pseudoscalar types
\begin{equation}
\label{L4q}
  {\cal L}_{NJL}=\frac{G}{2}\left[ (\bar{q}\lambda_aq)^2+
                           (\bar{q}i\gamma_5\lambda_aq)^2 \right]
\end{equation}
and the six-quark 't Hooft interaction 
\begin{equation}
\label{Ldet}
  {\cal L}_6=\kappa\, (\mbox{det}\ \bar{q}P_Lq
                         +\mbox{det}\ \bar{q}P_Rq),
\end{equation}
Here the positive coupling $G,\, [G]=\mbox{GeV}^{-2}$ has order $G\sim 1/N_c$,
the negative coupling $\kappa$,  $[\kappa]=\mbox{GeV}^{-5}$ with large 
$N_c$ asymptotics $\kappa\sim 1/N_c^{N_f}$.
The $P_{L,R}=(1\mp\gamma_5)/2$ are projectors and the 
determinant is over flavor indices. At large $N_c$,
${\cal L}_{NJL}$ dominates over ${\cal L}_6$.

This Lagrangian has been studied at mean field level in terms of quark degrees of freedom in \cite{Bernard:1988} 
 and in a bosonized form in \cite{Reinhardt:1988} and has been widely used since then \cite{Hatsuda:1994}.
.
\vspace{0.5cm}

{\bf Bosonization}
\vspace{0.5cm}

We use functional integral methods to bosonize the multi-quark Lagrangian, for which the
vacuum persistance amplitude is
\begin{equation}
\label{genf1}
   Z=\int {\cal D}q{\cal D}\bar{q}\,\exp\left(i\int d^4x\,{\cal L}\right).
\end{equation}
The fermion vertices of the original Lagrangian are linearized with help 
of the functional identity \cite{Reinhardt:1988}
\begin{eqnarray}
\label{1}
   1&\!\!\! =\!\!\!&\int \prod_a 
   {\cal D}s_a {\cal D}p_a
   \delta (s_a\color{black}-\bar{q}\lambda_aq)
       \delta (p_a\color{black}-\bar{q}i\gamma_5\lambda_aq)
       \nonumber \\
    &\!\!\! =\!\!\!&\int \prod_a {\cal D}s_a {\cal D}p_a
   {\cal D}\sigma_a {\cal
   D}\phi_a 
   \nonumber \\
   &\!\!\!\times\!\!\!&\,\exp\left\{i\int\! \ud^4 x
   \left[\sigma_a s_a
   -\bar{q}\lambda_aq) +
   \phi_a(p_a
   -\bar{q}i\gamma_5 \lambda_aq)\right]\right\}, \nonumber
\end{eqnarray}
 
\begin{eqnarray} 
\label{genf3}
   Z&\!\!=\!\!&\int\prod_a{\cal D}\sigma_a{\cal D}\phi_a
       {\cal D}q{\cal D}\bar{q}
       \exp\left(i\int d^4x {\cal L}_q(\bar{q},q,\sigma
        ,\phi\color{black})\right) 
       \nonumber \\
   &\!\!\times\!\!&\int \prod_a{\cal D}s_a{\cal D}p_a\,
       \exp\left(i\int d^4x\,{\cal L}_r(\sigma,
       \phi ,s,p)\right)
\end{eqnarray}
with
\begin{eqnarray}
\label{lagr2}
  {\cal L}_q(\bar{q},q,\sigma
  ,\phi)
  &=&\bar{q}(i\gamma^\mu\partial_\mu -\sigma 
     -i\gamma_5\phi )q, \\
\label{lagr3}
  {\cal L}_r(\sigma ,\phi
  ,s\color{black},p)
  &=&\frac{G}{2} \left( s_a^2
  +p_a^2 \right)
  +s_a(\sigma_a
  -{\hat m}_a) + p_a\phi_a \nonumber \\
  &+&\frac{\kappa}{32}\,A_{abc}\,s_a
  \left(s_bs_c-3\color{red}p_bp_c
  \right).
\end{eqnarray}
The totally symmetric constants $A_{abc}$ are defined 
through the flavor determinant
\begin{equation}
\label{flavA}
\det U = A_{abc} U_a U_b U_c\ ,
\end{equation}
where $U = U_a \lambda_a$.
The scalar fields $\sigma_a$, 
$a=0,3,8$ have non-zero vacuum expectation values
related to spontaneous breaking of global chiral symmetry, which 
give rise to the constituent masses $m_a$ of the quarks.
New scalar fields with zero vacuum expectation values
$\big<0|\sigma_a|0\big>=0$ must be defined through the shift 
$\sigma_a\rightarrow\sigma_a +m_a$. We seek the final bosonized Lagrangian expressed in
terms of tree level mesonic fields 
$\sigma_a$  and $\phi_a$ and must integrate out all remaining fields. 
\vspace{0.5cm}

{\bf Methods used for the functional integration}
\vspace{0.5cm}

The Gaussian integration over the quarks is done
using the 
generalized heat kernel technique developed in \cite{Osipov1:2001}.
It accounts for a chiral covariant treatment
of the one-loop determinant of the Dirac operator with an arbitrary 
non-degenerate quark mass matrix  at each order of an expansion with generalized Seeley-DeWitt 
coefficients.

The integral over the
auxiliary bosonic fields $s_a,p_a$ 
\begin{equation}
\label{pertI}
   I[\sigma_a
       ,\phi_a ] 
       = \int\prod_a{\cal D}s_a{\cal D}p_a
       \,
       \exp\left(i\int d^4x\,{\cal L}_r(\sigma,
       \phi ,s,p
       )\right).
\end{equation}
is done perturbatively. For that let us devide the lagrangian ${\cal L}_r$ in two parts. The free part, 
${\cal L}_0$, given by 

\begin{equation}
   {\cal L}_0 (\sigma ,\phi,s,p)
   =\frac{G}{2} \left(s_a^2
   +p_a^2\right)
   +s_a \sigma_a
   +p_a\phi_a. 
\end{equation}

The 't Hooft interaction is considered as perturbation ${\cal L}_I$ 

\begin{equation}
   {\cal L}_I (s_a,p_a) =   
   \frac{\kappa}{32}\,A_{abc}\,s_a
   \left(s_bs_c-3p_bp_c
   \right).
\end{equation}
For simplicity, a one-dimensional field theory approximation for 
the integral (\ref{pertI}) is considered. The extension to the real 
case is straightforward.
We put our system in the interval of size $L$, i.e. 
$-L/2\leq x\leq L/2$, assuming the limit $L\to\infty$ in the end 
of our calculations. 

The Fourier decomposition of the fields 
$f_a(x)=\{s_a, p_a, \sigma_a, \phi_a\}$ inside the interval is 

\begin{equation}
\label{fe}
     f_a(x)=\sum_{n=-\infty}^{+\infty} f_n^a\,
            \exp\left(i2\pi \frac{nx}{L}\right).
\end{equation}
This corresponds to the periodic boundary conditions 
$f_a(-L/2)=f_a(L/2)$. 
Then we have
\begin{equation}
     \int dx {\cal L}_r(x) \simeq 
     \int_{-L/2}^{L/2} dx\, {\cal L}_r(x) 
     =L\, {\cal L}_r\, ,
\end{equation}
where ${\cal L}_r={\cal L}_0+{\cal L}_I$, and  
\begin{equation} 
   {\cal L}_0=\sum_{n=-\infty}^{+\infty}\left[ 
   \frac{G}{2}\left(s^a_ns^a_{-n}+p^a_np^a_{-n}\right)
   +\sigma^a_n s^a_{-n} + \phi^a_n p^a_{-n}\right], 
\end{equation}
\begin{equation} 
   {\cal L}_I=\frac{\kappa}{32}A_{abc}\sum_{n,m,k}
   s^a_n \left( s^b_{m}s^c_k-3p^b_m p^c_{k}\right)\delta_{n+m+k,0}. 
\end{equation} 
The functional integral can be understood as the product of integrals
over the Fourier coefficients
\begin{equation}
     \int\limits_{-\infty}^{+\infty} {\cal D}s_a(x) {\cal D}p_a(x) \to
     \prod_{n} \int\limits_{-\infty}^{+\infty} ds_n^a dp_n^a.
\end{equation}
Thus, in the perturbative approximation we have
\begin{equation}
     I[\sigma ,\phi ]\sim 
     \exp \left\{iL \color{red} \hat {\cal L}_I\color{black}
     (S_n^a, P_n^a)\right\}
     \prod_{n} \int\limits_{-\infty}^{+\infty} ds_n^a dp_n^a
     \exp \left(iL \color{red} {\cal L}_0 \color{black}\right)
\end{equation} 
where
\begin{equation}
   S^a_n =-\frac{i}{L}\frac{\partial}{\partial \sigma^a_{-n}}\, ,\qquad 
   P^a_n =-\frac{i}{L}\frac{\partial}{\partial \phi^a_{-n}}\, . 
\end{equation}
It is convinient to use the normalized functional 
$I_N[\sigma ,\phi ]$ defined as
\begin{equation}
   I_N[\sigma ,\phi ] = \frac{I[\sigma , \phi ]}{I[0,0]}\, .
\end{equation}
The Gaussian functional integrals can be simply evaluated, for
instance, one has

\begin{eqnarray}
   I[\sigma ] 
   &\!\!\! =\!\!\!&
   \prod_n\int\limits_{-\infty}^{+\infty}ds_n \exp\left\{
   iL\sum_{n=-\infty}^{+\infty}
   \left(\frac{G}{2}s_ns_{-n}+\sigma_ns_{-n}\right)\right\}\nonumber\\
   &\!\!\! =\!\!\!& 
    I[0]\exp\left\{-\frac{iL}{2G}
   \sum_{n=-\infty}^{+\infty} \sigma_n\sigma_{-n}\right\}.
\end{eqnarray}
Thus the functional integration yields
\begin{eqnarray}
   I_N[\sigma ,\phi ] 
   &\!\!\sim\!\!&\, 
   \exp \left\{iL \hat {\cal L}_I
   (S_n^a, P_n^a)\right\}
   \nonumber \\
   &\!\!\times\!\!&\,\exp \left\{-\frac{iL}{2G} 
   \sum_{n=-\infty}^{+\infty} 
   \left(\sigma_n\sigma_{-n} +\phi_n\phi_{-n}\right)\right\}.
\end{eqnarray} 
One should calculate partial derivatives to obtain the effective
action $\Gamma_{eff}$
\begin{equation}
   I_N[\sigma ,\phi ]\sim 
   A[\sigma ,\phi ]\, e^{i\Gamma_{eff}}.
\end{equation}
We have
\begin{equation}
\label{cGamma_eff}
   \Gamma_{eff}
   =i\ln A[\sigma ,\phi ] +\Gamma_0 
   -
   i\ln\left(1+e^{-i\Gamma_0}
   \left(e^{i\hat{\Gamma}_I}-1\right)e^{i\Gamma_0}\right),
\end{equation}
where $\hat\Gamma_I=L \hat {\cal L}_I(S_n^a, P_n^a)$,
and $A[\sigma ,\phi ]$ is fixed by the requirement that the
effective action $\Gamma_{eff}$ is real. Here $\Gamma_0$ represents 
the leading order action.
\begin{equation}
\label{cgamma0}
   \Gamma_0 = - \frac{L}{2G}\sum_{n=-\infty}^{\infty}
   \left(\sigma^a_n\sigma^a_{-n}
   +\phi^a_n\phi^a_{-n}\right).
\end{equation}
The logarithm in eq.(\ref{cGamma_eff}) is a source of $U(1)$
breaking corrections which arise as a series in powers of the
partial derivatives. Expanding  


\begin{equation}
   \delta = e^{-i\Gamma_0}
   \left(e^{i\hat{\Gamma}_I}-1\right)e^{i\Gamma_0}
   = \sum_{n=1}^{\infty}\kappa^n\delta_n, 
\end{equation}
one can determine 
\begin{equation}
   \delta_1 = \frac{-iL}{32G^3}\, A_{abc}\sum_{n,m,k}\sigma^a_n 
   (\sigma^b_m\sigma^c_k - 3\phi^b_m\phi^c_k )\delta_{n+m+k,0}\, ,
\end{equation}
\begin{eqnarray}
   \delta_2 &\!\! = \!\!& \frac{\delta^2_1}{2}
   -i\,\frac{9L}{64(2G)^5}\, A_{abc}A_{a\bar{b}\bar{c}}
   \sum_{m,k,\bar{m},\bar{k}} \delta_{m+k+\bar{m}+\bar{k},0}
   \nonumber \\
   &\!\! \times\!\!& \left[\, 
   4\sigma^b_m\sigma^{\bar{b}}_{\bar{m}}\phi^c_k\phi^{\bar{c}}_{\bar{k}}
   + \left(\sigma^b_m\sigma^c_k - \phi^b_m\phi^c_k \right)
   \left(\sigma^{\bar{b}}_{\bar{m}}\sigma^{\bar{c}}_{\bar{k}} 
   - \phi^{\bar{b}}_{\bar{m}} \phi^{\bar{c}}_{\bar{k}} \right)\,\right]
   \nonumber \\  
   &\!\! +\!\!& 
   \frac{1}{(8G^2)^2}\sum_n \left( \sigma^a_n\sigma^a_{-n} +
   \phi^a_n\phi^a_{-n}\right)\sum_m 1
   \nonumber \\ 
   &\!\! +\!\!&\frac{3i}{32LG^3}\left(\sum_m 1\right)^2\, ,
\end{eqnarray}
and so forth.
The factor $A[\sigma ,\phi]$ must be also expanded

$$
   A[\sigma , \phi ] =
   1+\kappa\beta_1 +\kappa^2\beta_2 + {\cal O}(\kappa^3), 
$$

and, up to the considered order in $\kappa$, we have

$$
   \beta_1 =0, \quad 
   \beta_2 =\frac{1}{(8G^2)^2}\sum_n \left( \sigma^a_n\sigma^a_{-n} +
   \phi^a_n\phi^a_{-n}\right)\sum_m 1\, . 
$$

Therefore the perturbative action systematically obtains $U(1)$
breaking corrections in $\kappa^n$, which we collect in the
corresponding part of the action, $\Gamma_n$,
\begin{equation}
   \Gamma_{eff} = \sum_{n=0}^{+\infty} \Gamma_n\, ,
\end{equation}
where $\Gamma_0$ is given by eq.(\ref{cgamma0}), and
\begin{eqnarray}
\label{gammas}
   \Gamma_1 &\!\!\! =\!\!\!& -
   \frac{\kappa L}{32G^3}\, A_{abc}\sum_{n,m,k}\sigma^a_n 
   (\sigma^b_m\sigma^c_k - 3\phi^b_m\phi^c_k )\delta_{n+m+k,0}\, ,
   \nonumber\\
   \Gamma_2 &\!\!\! =\!\!\!&
   -\frac{9\kappa^2 L}{64(2G)^5}\, A_{abc}A_{a\bar{b}\bar{c}}
   \sum_{m,k,\bar{m},\bar{k}} \delta_{m+k+\bar{m}+\bar{k},0}\left[\, 
   4\sigma^b_m\sigma^{\bar{b}}_{\bar{m}}\phi^c_k\phi^{\bar{c}}_{\bar{k}}
   \right. \nonumber \\
   &\!\!\! +\!\!\!&\left.
   \left(\sigma^b_m\sigma^c_k - \phi^b_m\phi^c_k \right)
   \left(\sigma^{\bar{b}}_{\bar{m}}\sigma^{\bar{c}}_{\bar{k}} 
   - \phi^{\bar{b}}_{\bar{m}} \phi^{\bar{c}}_{\bar{k}}
   \right)\,\right]. 
\end{eqnarray}
Here an unessential constant has been omitted in $\Gamma_2$.
Taking the infinite -- volume limit $L\to\infty$, one finally obtains

$$
   \Gamma_0 = -\frac{1}{2G}\int\limits_{-\infty}^{+\infty} dx\,
   \left(\sigma^2_a(x)+\phi^2_a(x)\right)\, ,
$$

$$
   \Gamma_1 = 
   - \frac{\kappa}{32G^3}\, A_{abc}
   \int\limits_{-\infty}^{+\infty}\! dx\, \sigma_a(x)\left[ 
   \sigma_b(x)\sigma_c(x) - 3\phi_b(x)\phi_c(x)\right],
$$

$$
   \Gamma_2 = 
   -\frac{9\kappa^2}{64(2G)^5}\, A_{abc}A_{a\bar{b}\bar{c}}
   \int\limits_{-\infty}^{+\infty}\! dx\, \left\{\, 
   4\sigma_b(x)\sigma_{\bar{b}}(x)\phi_c(x)\phi_{\bar{c}}(x)
   \right. 
$$

$$
   +\left.
   \left[ \sigma_b(x)\sigma_c(x) - \phi_b(x)\phi_c(x) \right]
   \left[ \sigma_{\bar{b}}(x)\sigma_{\bar{c}}(x) 
   - \phi_{\bar{b}}(x) \phi_{\bar{c}}(x)
   \right]\,\right\},
$$

and so forth.
The infinite volume limit for the coefficient $\beta_2$ yields 

$$
\label{beta2}
   \beta_2 =\frac{1}{(8G^2)^2} \int_{-L/2}^{L/2}\! dx\, 
   \left( \sigma_a(x)\sigma_a(x) +
   \phi_a(x)\phi_a(x) \right) \frac{1}{L}
   \sum_{m=-\infty}^{+\infty} 1\, . 
$$

The integral has a smooth $L\rightarrow\infty$ limit, but contains 
a local factor, which can be understood as $\delta$-function 
singularity,  $\delta (0)$, see also \cite{Osipov:2002}, \cite{Osipov:2004}.
To show this one needs the following result
\begin{equation}
\label{deltasum}
     \sum_{n=-\infty}^{+\infty}\exp\left(i\frac{2\pi x}{L}n\right)
     =L\sum_{n=-\infty}^{\infty}\delta (x-Ln)=L\delta (x).
\end{equation}

On the last step we took into account that in the considered
problem all $x$-dependent functions are integrated only in the 
interval $-L/2\leq x\leq L/2$ and, therefore, only the term with 
$n=0$ contributes. The infinite value
\begin{equation}   
   \frac{1}{L}\sum_{n=-\infty}^{+\infty}1
   = \delta (0)  
\end{equation}        
represents the density of Fourier harmonics in the interval. It 
must be regularized by cutting an upper part of the spectrum, e.g.
\begin{equation}   
     \delta (0)_{\mbox{\footnotesize reg}}
     = \frac{1}{L}\sum_{n=-N}^{N}1 = \frac{2N+1}{L}
\end{equation}       
where $N$ is large enough. 
One can fix $N\gg 1$ without any relation to the size of the box.
In this case the limit $L\to\infty$ leads to the vanishing of the $\delta $-function and to $\beta_2=0$.
Alternatively, one can relate $N$ with $L$ by introducing the 
momentum space cutoff $\Lambda_E$: $N(L)=L\Lambda_E/(4\pi)\gg 1$. 
Unlike $L$ the cutoff $\Lambda_E$ has an obvious physical meaning 
giving the scale of momenta relevant for the problem. 
Indeed, the $n$th harmonic has a momentum $p_n=2\pi n/L$. 
The size of the considered box is the difference 
$p_N-p_{-N}=4\pi N/L=\Lambda_E$. This scale cannot be eliminated 
by taking the limit $L\to\infty$. One has instead
\begin{equation}
     \delta (0)_{\mbox{\footnotesize reg}}=\frac{\Lambda_E}{2\pi}
     +\frac{1}{L}\, ,       
\end{equation}     
where only the second term does not contribute in the limit 
$L\rightarrow\infty$. By introducing the cutoff $\Lambda_E$, 
we suppose that the density of Fourier harmonics has a 
finite value which can be fixed phenomenologically. In this case $\beta_2\neq 0$. 
It is a pure physical matter of argument to choose
among these two alternatives.

The Reinhardt and Alkofer estimate \cite{Reinhardt:1988}
takes the integral (\ref{pertI}) at one stationary phase point    
without any integration over the auxiliary variables $s_a, p_a$.
 
The contributions at the stationary phase condition are obtained as 
\begin{eqnarray}
\label{Lr}
  {\cal L}_r(r_{\mbox{st}})&\!\! =\!\!& 
  \frac{G}{12}\ \mbox{tr}\ (U_{st} U^\dagger_{st} )
             + \frac{1}{6}\ \mbox{tr}\ (WU_{st}^\dagger + W^\dagger
               U_{st} ) \nonumber \\
  &\!\!=\!\!& h_a\sigma_a 
             + \frac{1}{2}h^{(1)}_{ab}\sigma_a\sigma_b
             + \frac{1}{2}h^{(2)}_{ab}\phi_a\phi_b + \ldots\ .
\end{eqnarray}
fulfilling
\begin{equation}
\label{spc}
   GU_a + W_a + \frac{3\kappa}{32}A_{abc}U^\dagger_bU^\dagger_c = 0
\end{equation}
with covariant combinations 
$$
   W_a=\sigma_a +\Delta_a -i\phi_a,
$$ 
and definitions 
$$
   \Delta_a=m_a-{\hat m}_a, \qquad  U_a = s_a -ip_a.
$$
The field $U_{st}$ represents the exact 
solution of condition (\ref{spc}), obtained by expanding $s_a, p_a$ 
in  parity even and odd combinations 
respectively, of increasing powers of bosonic fields $\phi_a, \sigma_a$

\begin{eqnarray}
\label{rsta}
       s^a_{st}
       &\!\! =\!\!& h_a+h_{ab}^{(1)}\sigma_b
       +h_{abc}^{(1)}\sigma_b\sigma_c
       +h_{abc}^{(2)}\phi_b\phi_c 
       \nonumber \\
       &\!\! +\!\!& h_{abcd}^{(1)}\sigma_b\sigma_c\sigma_d
       +h_{abcd}^{(2)}\sigma_b\phi_c\phi_d
       +\ldots \nonumber \\
       p^a_{st}
       &\!\! =\!\!& h_{ab}^{(2)}\phi_b
       +h_{abc}^{(3)}\phi_b\sigma_c
       +h_{abcd}^{(3)}\sigma_b\sigma_c\phi_d
       \nonumber \\
     &\!\! +\!\!&h_{abcd}^{(4)}\phi_b\phi_c\phi_d
       +\ldots. 
\end{eqnarray}
The coefficients $h^{(k)}_{ab...}$ 
depend explicitly on the quark 
masses and coupling constants $G,\kappa$ and  are fixed by
a series of coupled equations following from (\ref{spc}) and
obtained by equating to zero the factors before independent 
combinations of mesonic fields. 
Due to recurrency of the considered 
equations all coefficients are determined once the first one, 
$h_a$, has been obtained. 
Note that the stationary phase condition (\ref{spc}) has several 
solutions \cite{Osipov:2004}. If one wants to apply the stationary phase (SP) method 
consistently, one must take into account the contributions of all 
critical points (even at leading order). 
The $1/N_c$ counting indicates that the last term in 
the stationary phase eq.(\ref{spc}) is $1/N_c$ suppressed. This
argument can be used for the systematic $1/N_c$ expansion of the 
effective action, what actually has been done in our perturbative
approach. 

We have found strong indications that the estimate \cite{Reinhardt:1988} 
corresponds to a resummation
of the perturbative series in $\kappa$, collecting all tree (non-singular) contributions related with the auxiliary fields $s_a,p_a$ of the perturbative approach. Further terms of the ressummed series yield systematically "loop"-corrections of these fields \cite{Osipov:5}.
\vspace{0.5cm}

{\bf Results and Discussion}
\vspace{0.5cm}

We refer to \cite{Osipov:2005b} for a detailed derivation and calculation of the leading order low lying pseudoscalar and scalar meson spectra and related observables. Further sets are also presented in \cite{Osipov:2005}. Already in its simplest form, the Lagrangian yields overall quite good fits, even the scalars are reasonable. They have been obtained with the following approximations.

a) Related with the evaluation of the Gausiian functional integral over quarks: 
knowing that the lack of confinement
in the NJL model introduces serious difficulties with the crossing of
non-physical thresholds associated with the production of free 
quark - antiquark pairs, which one may encounter by formally 
continuing the full Euclidean action to Minkowski space,
the heat kernel series has been truncated at second order.
We admittedly deviate from the original NJL Lagrangian, however in a way which 
relies heavily on its symmetries and asymptotic dynamics, which are 
fully taken into account. 
  
b) Related with the functional integral calculation over the auxiliary variables $s_a,p_a$, of cubic order:
we have used the SPA with one critical point. We know however that there exist more critical points which should be taken into account even in leading order in SPA. 
\vspace{0.5cm}

Finally we have suggested a different way of bosonization, considering the ´t Hooft determinant as a perturbation to the four-quark intercation. One can relate the perturbative expansion to the SPA with one critical point through a resummation. This is discussed in full detail in \cite{Osipov:5}. Corrections to the leading order can be obtained in a systematic way.
Due to the local character of the interactions these corrections involve a divergent density of harmonics. We discuss how a cutoff can be introduced, and together with the ultraviolet cutoff related with the quark loops, we obtain an effective bosonized Lagrangian. Calculation of the related mass spectra is in progress \cite{Osipov:6}. 
\vspace{0.5cm}

{\bf Acknowledgements}
We are very grateful to the organizers for the invitation and for hosting a very interesting and enjoyable meeting.
This work has been supported by grants provided by Funda\c c\~ao para
a Ci\^encia e a Tecnologia, POCTI/35304/FIS/2000, POCTI/FNU/50336/2003 and Centro de F\ii sica
Te\'orica unit 535/98. 
This research is part of the EU integrated infrastructure initiative 
HadronPhysics project under contract No.RII3-CT-2004-506078.
A.A. Osipov also gratefully acknowledges the  
Funda\ca o Calouste Gulbenkian for financial support.

\end{document}